\documentclass[a4paper]{article}
\usepackage{mn_wjp}
\usepackage{graphicx}
\def\msun{{\rm\,M_\odot}}

\begin{document}

\title[Cosmological Evolution] 
{Cosmological evolution and hierarchical galaxy formation}
\author[W.Percival and L.Miller]{Will Percival and Lance Miller \\
Dept. of Physics, University of Oxford, 
Nuclear \& Astrophysics Laboratory,
Keble Road, Oxford OX1 3RH, U.K.\\}

\date{Received in original form  }

\maketitle

\begin{abstract} 
We calculate the rate at which dark matter halos merge to form higher
mass systems. Two complementary derivations using Press-Schechter
theory are given, both of which result in the same equation for the
formation rate. First, a derivation using the properties of the
Brownian random walks within the framework of Press-Schechter theory
is presented.  We then use Bayes' theorem to obtain the same result
from the standard Press-Schechter mass function. The rate obtained is
shown to be in good agreement with results from Monte-Carlo and N-body
simulations. We illustrate the usefulness of this formula by
calculating the expected cosmological evolution in the rate of star
formation that is due to short-lived, merger-induced starbursts. The
calculated evolution is well-matched to the observed evolution in
ultraviolet luminosity density, in contrast to the lower rates of
evolution that are derived from semi-analytic models that do not
include a dominant contribution from starbursts. Hence we suggest that
the bulk of the observed ultraviolet starlight at $z > 1$ arises from
short-lived, merger-induced starbursts. Finally, we show that a simple
merging-halo model can also account for the bulk of the observed
evolution in the comoving quasar space density.
\end{abstract}

\begin{keywords}
galaxies: formation, galaxies: starburst, galaxies: active, cosmology: theory
\end{keywords}

\section{Introduction}

If the matter content of the universe is dominated by cold dark
matter, then dark matter halos associated with galaxies are expected
to form hierarchically.  Bond et al. \shortcite{bond} have shown that
a more rigorous treatment of the work of Press \& Schechter
\shortcite{ps} can be used to obtain information about the build-up of
structure.  Calculations using Press-Schechter (PS) theory are based
on a statistical analysis of the initial field of density
perturbations and do not include any non-linear dynamical effects, but
nonetheless the results appear to agree well with dynamical,
collisionless (``N-body'') simulations \cite{lc94,somerville}.

The aim of this paper is to use PS theory to calculate the rate of
formation of dark halos of specified mass.  We compare the results
with those of Monte-Carlo and N-body simulations, and we also use the
calculations to argue that the bulk of the observed cosmological
evolution in star-formation and quasar activity is a reflection of the
evolution in rate of dark halo formation.

A standard application of PS theory is to calculate the mass function
given a cosmic epoch of interest. In many cosmological situations we
are also interested in the rate at which halos of some mass form.
Consequently, there have been a number of attempts to estimate this
rate using PS theory~\cite{lc93,sasaki}.  In this paper we use the
theory of random walks within the PS framework to calculate directly
the halo formation rate for any cosmology.  We also use Bayes' theorem
to derive the same equation from the standard PS mass function. These
complementary methods provide important insights into how the
formation of halos can be understood within PS theory.

In section~\ref{sec:mc} we use a Monte-Carlo realisation of actual
Brownian random walks to calculate the formation epochs of the halos
they represent. The distribution of these epochs is found to be in
good agreement with the formula calculated previously. Following this
we show that the formation time distribution of halos in a large
N-body simulation (found using a standard friend-of-friends algorithm)
is also well approximated by the PS result.

We then discuss the relevance of this work to our understanding of the
observed cosmological evolution in star-formation rate (SFR) and
quasar space density.  In particular we suppose that there is a causal
link between the hierarchical formation of dark halos and the
activation of both quasars and luminous bursts of star formation.  The
possibility of a connection between these two key evolving quantities
has already been suggested by a number of authors
\cite{shaver,dunlop,bt}. Mergers and interactions have long been
implicated in both luminous starbursts and active galaxies (see the
review by Barnes \& Hernquist 1992 and refs therein) and merging
galaxies have been found in which {\em both} phenomena are observed
\cite{canalizo,stockton,brotherton}.  Investigation of the galaxies
found at redshifts around 3 by the Lyman-break method \cite{steidel97}
show that much of the ultraviolet light inferred to be due to star
formation can be identified with individual galaxies with star
formation rates of possibly up to $1000\msun$ per year
\cite{steidelrs}.  These galaxies are strongly clustered and hence are
likely to be the progenitors of cluster galaxies at the present epoch
(Steidel et al.\ 1998a,b; Adelberger et al.\ 1998): at high redshifts
there is therefore direct evidence for a link between the high level
of star formation and host galaxies which are inferred to be merging
into higher-mass galaxy systems.  The detailed physics of both quasar-
and star-formation is complicated and not understood, and in this
paper we discuss only the role that the cosmological variation in halo
formation rate might have in determining the evolution in observed
quantities such as the ultraviolet luminosity density arising from
star formation.

\section{Press-Schechter theory} \label{sec:ps}

We now describe briefly the principles of Press-Schechter (PS) theory
and calculate the number density of halos at a particular epoch - the
standard PS formula; a result which has been shown to be in good
agreement with N-body simulations (e.g.\ Efstathiou et al.\ 1988).

Dark halos are assumed to form by the non-linear gravitational
collapse of initial density perturbations.  We assume that the initial
perturbations form a homogeneous, isotropic Gaussian random field.  In
PS theory such a field is smoothed by convolving with a filter
function whose size is related to the mass $M$ of halo in which we are
interested.  The fractional overdensity at any location is assumed to
grow linearly until a critical overdensity, $\delta_c$, is reached,
when that location is considered to have collapsed into a dark halo of
mass $M$, provided that the critical overdensity is not exceeded when
the field is filtered on a scale corresponding to a larger mass.  In
the analysis that follows, instead of viewing the field as growing
with time, we consider the field to be fixed and the critical
overdensity $\delta_c$ to decrease with time.

If we use a sharp k-space filter of radius R,
$W(\mathbf{r};R)=(\sin(r/R)-(r/R)\cos (r/R))/(r/R)^{3}$, the density
of the filtered field at any point is given by a Brownian random walk
with the variance of the filtered field being the `time' axis
\cite{peacock,bond}. The mass associated with this point in space is
given by the position of the first upcrossing of an absorbing barrier
at $\delta=\delta_c(t)$. The probability density function that a
trajectory will have its first upcrossing at a mass between $M$ and
$M+dM$, $P(M|\delta_c)dM$ is given by a solution of the diffusion
equation with an absorbing boundary condition \cite{bond}:
\begin{equation}
   P(M|\delta_c)dM=\frac{\delta_c}
      {(2\pi)^{1/2}\sigma_{M}^{3}}
      \exp\left(-\frac{\delta_c^{2}}{2\sigma_M^2}\right)
      \left|\frac{d\sigma_M^2}{dM}\right|dM \label{eq:psprob}
\end{equation}
where $\sigma_M^2$ is the variance of the filtered field which for
a sharp k-space filter is given by:
\begin{equation}
  \sigma_M^2=\frac{1}{2\pi^{2}}\int_{0}^{2/R}k^{2}P(k)dk \label{eq:sigsq}
\end{equation}
where $P(k)$ is the power spectrum of the initial random field. We can
use equation~\ref{eq:psprob} to obtain the comoving number density of
halos of mass M at a given time t, $n(M|t)$:
\begin{displaymath}
  n(M|t)MdM=\rho P(M|\delta_c(t))dM
\end{displaymath}
\begin{equation}
  \hspace{1cm}
  =2\rho\frac{\delta_c(t)}{(2\pi)^{1/2}\sigma_M^2}\exp\left(
  -\frac{\delta_c(t)^{2}}{2\sigma_M^2}\right)\left|\frac{d\sigma_{M}}{dM}
  \right|dM.  \label{eq:ps}
\end{equation}
This is the standard PS mass function. It should be noted that when
using other filters this result is not correct
\cite{bond,jedamsik,yano}.

To calculate the numbers of halos and their rate of formation we need
to normalise the power spectrum.  We require the variance in the
density field when filtered with a top-hat filter of radius
$8h^{-1}$\,Mpc, $\sigma_8$, to match the values deduced from X-ray
clusters \cite{eke}.

For a top-hat filter, we can relate the mass of a halo to the filter
radius, $M=4/3\pi{\rho}R^{3}$. We can also calculate the critical
overdensity for collapse for a uniform spherical region; for a flat
$\Omega_{\Lambda}=0$ universe this is given by $\delta_c=\alpha(1+z)$,
where $\alpha\sim1.68$ \cite{gunn}. For an open $\Omega_{\Lambda}=0$
universe, $\delta_c(z)$ is given by Lacey \& Cole \shortcite{lc93},
and for a flat $\Omega_{\Lambda}\neq0$ universe by Eke, Cole \& Frenk
\shortcite{eke}.  For a sharp k-space filter, the relationship between
the mass and filter size is less obvious.  Following Lacey and Cole
\shortcite{lc93}, we integrate the filter function over all space to
obtain its `volume of influence', which gives $M=3/4\pi^{2}\rho
R^{3}$.  However, if we use the critical overdensity applicable to the
top-hat filter used to normalise the power spectrum the k-space filter
predicts a different number density of halos.  We therefore choose the
critical overdensity associated with a sharp k-space filter to be that
which predicts the same number density of halos for both filters at
the mass used to normalise the power spectrum.

\section{Derivation of the halo formation rate from PS theory} \label{sec:der1}

We now use PS theory to obtain more information about the build-up of
structure, in particular to calculate the {\em rate} at which dark
halos of a particular mass form. We cannot calculate this simply by
taking the time derivative of equation~\ref{eq:ps} as, in PS theory,
halos are not only being continually formed at any particular mass but
are also continually being lost into halos of higher mass. We present
new work to calculate the required halo formation rate by examining
properties of the Brownian random walks invoked in standard PS
theory. We use the fact that the rate of halo formation is equivalent
to a probability density function in time. The analysis of Brownian
random walks is important in many fields of pure and applied science
and similar results to those derived below can also be found in many
introductory books on stochastic processes (e.g. Karlin \& Taylor
1975).

Initially, suppose that we are not interested in how the halo formed
so we are not interested in the shape of the trajectory up until the
epoch of halo formation. We wish to calculate the distribution of
cosmic epochs at which halos of a given mass $M$ are first formed,
$P(t|M)$dt. In PS theory, this probability density function is related
by a variable transformation to $P(\delta_c|\sigma_M^2)$, the
probability that a trajectory has its first upcrossing between
$\delta_c$ and $\delta_c+d\delta_c$ at a given value of $\sigma_M^2$,
and it is a formula for this that we now derive using the trajectories
approach. We let $\delta(\sigma_M^2)$ be the position of the walk at
$\sigma_M^2$ and define:

\begin{equation}
  \delta_{\rm max}(\sigma_M^2)=\mbox{\rm max}\{\delta(\sigma_{M'}^2);
    0\leq\sigma_{M'}^2\leq\sigma_M^2\},
\end{equation}
\begin{equation}
  \delta_{\rm diff}(\sigma_M^2)=\delta_{\rm max}(\sigma_M^2)-\delta(\sigma_M^2).
\end{equation}
\begin{figure}
  \centering \resizebox{7cm}{9.5cm}{\includegraphics{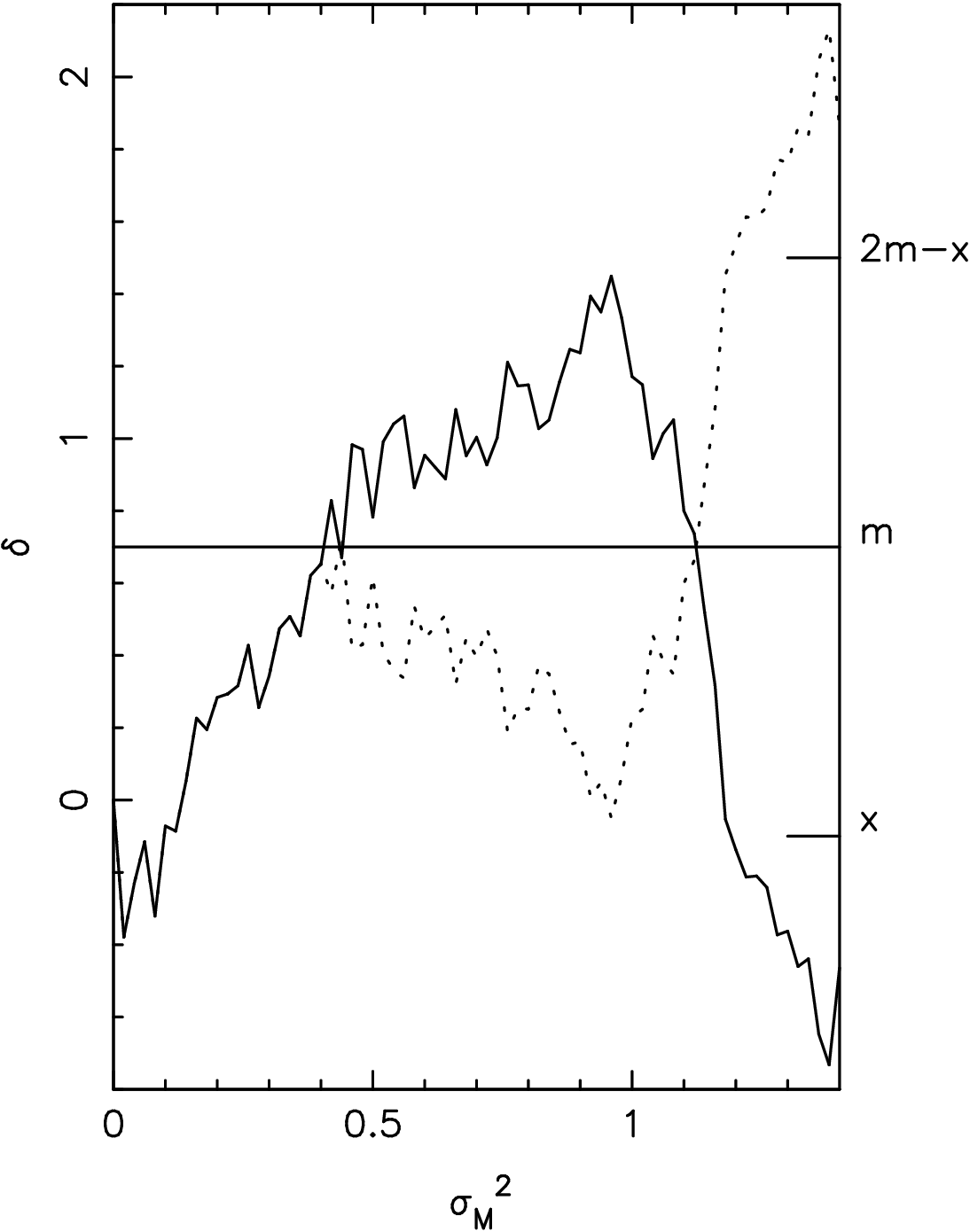}}
  \caption{The solid line represents a random walk consisting of 70
  equally spaced steps of $\sigma_M^2$ in the interval
  $0<\sigma_M^2<1.4$. The random walk was chosen from a set of such
  walks and is such that $\delta_{\rm max}(1.4)>m$ and
  $\delta(1.4)<x$. The dotted line represents the walk obtained by
  reflecting the initial trajectory about the line $\delta=m$ after
  the first upcrossing of this line. Note that it has the property
  that $\delta(1.4)>2m-x$. Similarly, any trajectory which is such
  that $\delta(1.4)>2m-x$ can undergo a similar reflection to obtain a
  trajectory with $\delta_{\rm max}(1.4)>m$ and $\delta(1.4)<x$. There is 
  a one-to-one correspondence between walks with
  $\delta_{\rm max}(\sigma_M^2)>m$ and $\delta(\sigma_M^2)<x$ and walks
  with $\delta(\sigma_M^2)>2m-x$. }
\label{fig:walk}
\end{figure}
By considering the reflection of trajectories about the line
$\delta=m$ beyond the first upcrossing of this line, we can see that
there is a one-to-one correspondence between trajectories for which
$\delta(\sigma_M^2)>2m-x$ and trajectories for which
$\delta_{\rm max}(\sigma_M^2)>m$ and $\delta(\sigma_M^2)<x$ where
$x<m$. For a pictorial representation of this correspondence see
Fig.~\ref{fig:walk}. This means that:
\begin{displaymath}
  P(\delta_{\rm max}(\sigma_M^2)\geq m, \delta(\sigma_M^2)\leq x |\sigma_M^2)
\end{displaymath}
\begin{displaymath}
  \hspace{1cm}
    = P(\delta(\sigma_M^2)\geq 2m-x|\sigma_M^2)
\end{displaymath}
\begin{equation}
  \hspace{1cm}
  = \frac{1}{2}\left[1-\mbox{erf}\left(\frac{2m-x}
    {(2\sigma_M^2)^{1/2}}\right)\right],
    m\geq0, m\geq x.
\end{equation}
We differentiate with respect to x and then with respect to m,
changing the sign, to get the joint density function for
$\delta_{\rm max}(\sigma_M^2)$ lying in the interval $m,m+dm$ and
$\delta(\sigma_M^2)$ lying in the interval $x,x+dx$. This gives:
\begin{displaymath}
  f(m,x|\sigma_M^2)dmdx
\end{displaymath}
\begin{equation}
    \hspace{1cm}
    =\sqrt{\frac{2}{\pi}}\frac{2m-x}{(\sigma_M^2)^{3/2}}
    \exp\left(\frac{-(2m-x)^2}{2\sigma_M^2}\right)dmdx,
\end{equation}
\begin{displaymath}
    \hspace{1cm} 0\leq m, x\leq m.
\end{displaymath}
To obtain the joint density function for $\delta_{\rm max}(\sigma_M^2)$
lying in the interval $m,m+dm$ and $\delta_{\rm diff}(\sigma_M^2)$ lying
in the interval $y,y+dy$ we note that:
\begin{displaymath}
  P(\delta_{\rm max}(\sigma_M^2)\leq a, \delta_{\rm diff}(\sigma_M^2)\leq b|\sigma_M^2)
\end{displaymath}
\begin{equation}
    \hspace{1cm}
    = \int_{0}^{a}\int_{0}^{b}f(m,m-y)dydm.
\end{equation}
From this we can deduce that the desired joint density is g(m,y)=f(m,m-y):
\begin{displaymath}
  g(m,y|\sigma_M^2)dmdy
\end{displaymath}
\begin{equation}
  \hspace{1cm}
    =\sqrt{\frac{2}{\pi}}\frac{m+y}{(\sigma_M^2)^{3/2}}
    \exp\left(\frac{-(m+y)^2}{2\sigma_M^2}\right)dmdy,
  \label{eq:myjoint}
\end{equation}
\begin{displaymath}
    \hspace{1cm} m\geq 0, y\geq 0.
\end{displaymath}
We now calculate $g(m|y=0,\sigma_M^2)$ using Bayes' theorem to
transform from equation~\ref{eq:myjoint} to the conditional probability
required and then setting $y=0$. This is equivalent to setting $y=0$
in equation~\ref{eq:myjoint} and renormalising so the probability
density function integrates to 1. Integrating $g(m,y|\sigma_M^2)dmdy$
we see that:
\begin{equation}
  g(y|\sigma_M^2)dy=\sqrt{\frac{2}{\pi}}\frac{1}{(\sigma_M^2)^{1/2}}
    \exp\left(\frac{-y^2}{2\sigma_M^2}\right)dy,y\geq 0,
\end{equation}
and applying Bayes' theorem we find that:
\begin{equation}
  g(m|y,\sigma_M^2)=\frac{m+y}{\sigma_M^2}
    \exp\left(\frac{-(m+y)^2+y^2}{2\sigma_M^2}\right)dm,
  \label{eq:mgiveny}
\end{equation}
\begin{displaymath}
  \hspace{1cm}m\geq 0.
\end{displaymath}
It just remains to set $y=0$ and change the variable from m to
$\delta_c$ which we now take to be the position of the first
upcrossing:
\begin{equation}
  P(\delta_c|\sigma_M^2)d\delta_c=\frac{\delta_c}{\sigma_M^2}
    \exp\left(-\frac{\delta_c^{2}}{2\sigma_M^2}\right)d\delta_c.
\end{equation}
This is the probability that a trajectory has its first upcrossing
between $\delta_c$ and $\delta_c+d\delta_c$ at a given value of
$\sigma_M^2$.

By changing the variables from $\sigma_M^2$ to mass $M$ and from
$\delta_c$ to time $t$, we obtain the probability that a halo of
mass $M$ formed between times $t$ and $t+dt$.  This gives the
dependence on cosmic epoch of the rate of halo formation:
\begin{equation}
  P(t|M)dt=\frac{\delta_c}{\sigma_M^2}
    \exp\left(-\frac{\delta_c^{2}}{2\sigma_M^2}\right)
    \left|\frac{d\delta_c}{dt}\right|dt. \label{eq:rate_cstm}
\end{equation}

In the special case of a flat $\Omega_{\Lambda}=0$ cosmology for which
$\delta_c=\alpha(1+z)$, this time evolution is given by the simple
formula:
\begin{equation}
  \left(\frac{dn}{dt}\right)^{+}\propto(1+z)^{3.5}\exp(-\beta(1+z)^{2})
  \label{eq:simple}
\end{equation} 
where $\beta=\alpha^{2}/(2\sigma_M^2)$ is a function of the mass
of halo and the power spectrum.

We note that because the trajectories are Brownian random walks, all
walks which pass through a given point can be thought of as new walks
starting from that point. It is thus possible to apply coordinate
changes to equation~\ref{eq:psprob} and obtain the conditional
probability for the mass distribution of progenitors of halos
\cite{bond}. Using Bayes' theorem and this result it is also possible to
calculate the distribution of final collapsed masses attained by
halos given the mass at an earlier time \cite{lc93}.

We can apply a similar argument to the result derived above for the
time distribution of formation events and obtain the distribution of
times at which a halo formed, given that it formed
part of a larger halo at a known later time. If we
consider walks starting from ($\sigma_{M'}^{2},\delta_c'$),
where $\sigma_{M'}^{2}<\sigma_M^2$ and $\delta_c'<\delta_c$
(i.e. $M'>M$ and $t'>t$) we obtain the required conditional
probability:
\begin{displaymath}
  P(\delta_c|\sigma_M^2,\delta_c',\sigma_{M'}^{2})d\delta_c
\end{displaymath}
\begin{equation} 
  \hspace{1cm}
  =\frac{\delta_c-\delta_c'}{\sigma_M^2-\sigma_{M'}^{2}}
  \exp\left(-\frac{(\delta_c-\delta_c')^{2}}
    {2(\sigma_M^2-\sigma_{M'}^{2})}\right)d\delta_c.
  \label{eq:condt}
\end{equation}

The behaviour of a trajectory, having passed through a point related
to the formation of a halo, thus provides mass and formation time
distributions for the progenitors of the halo. As noted above, the
form of the trajectory continuing from such a formation point is
independent of the position of that point, and consequently the mass
distribution of progenitors immediately prior to the formation event
is independent of the formation epoch. Turning this argument around,
we see that the effect of placing constraints on the progenitors of
the halo immediately prior to the formation event doesn't affect the
distribution of formation times. We thus have the important result
that, even if we only consider formation events resulting from mergers
with specified progenitors, such as halos resulting from a merger of
two approximately equally sized objects, the distribution of times at
which these mergers occur is still that given by
equation~\ref{eq:rate_cstm}.  It is for this reason that we consider
the cosmic variation in halo formation rate also to be a measure of
the cosmic variation in the rate of mergers between dark halos.  We
should note that because standard PS theory does not account for halo
sub-structure, the derivation presented here excludes possible mergers
between low-mass halo sub-units forming part of a larger collapsed
halo.

\section{An alternative derivation} \label{sec:der2}

We will now show that $P(t|M)$dt as given by
equation~\ref{eq:rate_cstm} can be calculated from the conditional
probability density function $P(M|t)$dM \cite{bond} using Bayes'
theorem.  Firstly, we wish to calculate the prior for $\delta_c$ which
is equivalent to asking the question `Given no knowledge of
$\sigma_M^2$, what is the distribution of upcrossing points in
$\delta_c$?'. We note that all trajectories must have an upcrossing
point of any given line $\delta=\delta_c$.  Now, in any two equally
sized intervals, $d\delta_1$ and $d\delta_2$, there must be equal
probability of such a crossing existing because the walk does not
alter its form at different $\delta_c$ (all walks which pass through a
given point can be thought of as new walks starting from that
point). Thus, given no {\em a priori} information about $\sigma_M^2$,
all values of $\delta_c$ are equally likely and we should assume a
uniform prior for $\delta_c$. In order to be mathematically rigorous,
we must make $\delta_c$ bounded. We see later that we can remove these
bounds without affecting the result. So we have that:
\begin{equation}
  P(\delta_c)d\delta_c=\left\{ \begin{array}{ll}
    \frac{d\delta_c}{\delta_{\rm max}-\delta_{\rm min}} &
    \mbox{$\delta_{\rm min}\leq\delta_c\leq\delta_{\rm max}$} \\ 0 &
    \mbox{otherwise.} \end{array} \right.
\end{equation}
Applying Bayes' theorem and using equation~\ref{eq:psprob}, we can
find the joint probability of $\delta_c$ and $\sigma_M^2$:
\begin{displaymath}
  P(\sigma_M^2,\delta_c)d\sigma_M^2d\delta_c=
  \frac{1}{(\delta_{\rm max}-\delta_{\rm min})}
\end{displaymath}
\begin{equation}
  \hspace{1cm}
  \times\frac{\delta_c}{(2\pi)^{1/2}(\sigma_M^2)^{3/2}}
    \exp\left(-\frac{\delta_c^{2}}{2\sigma_M^2}\right)
    d\sigma_M^2d\delta_c. \label{eq:rate_joint}
\end{equation}
We can integrate this equation over all $\delta_c$ to obtain the
probability density function for $\sigma_M^2$: 
\begin{displaymath}
  P(\sigma_M^2)d\sigma_M^2 =
  \frac{1}{(2\pi)^{1/2}(\sigma_M^2)^{1/2}(\delta_{\rm max}-\delta_{\rm min})}
\end{displaymath}
\begin{equation}
  \hspace{1cm}
  \times\left[\exp\left(-\frac{\delta_{\rm min}^{2}}{2\sigma_M^2}\right)
  - \exp\left(-\frac{\delta_{\rm max}^{2}}{2\sigma_M^2}\right) \right],
\end{equation}
and again use Bayes' theorem to obtain the conditional probability
$P(\delta_c|\sigma_M^2)$:
\begin{displaymath}
  P(\delta_c|\sigma_M^2)d\delta_c=\frac{\delta_c}{\sigma_M^2}
    \exp\left(-\frac{\delta_c^{2}}{2\sigma_M^2}\right)d\delta_c
\end{displaymath}
\begin{equation}
  \hspace{1cm}
  \times\left[\exp\left(-\frac{\delta_{\rm min}^{2}}{2\sigma_M^2}\right)
  - \exp\left(-\frac{\delta_{\rm max}^{2}}{2\sigma_M^2}\right) \right].
\end{equation}
In the limit as $\delta_{\rm min}\rightarrow0$ and
$\delta_{\rm max}\rightarrow\infty$ this becomes:
\begin{equation}
  P(\delta_c|\sigma_M^2)d\delta_c=\frac{\delta_c}{\sigma_M^2}
    \exp\left(-\frac{\delta_c^{2}}{2\sigma_M^2}\right)d\delta_c.
\end{equation}
This is the probability that given a particular value of $\sigma_M^2$,
the first upcrossing at this $\sigma_M^2$ has probability
$P(\delta_c|\sigma_M^2)$ of being between $\delta_c$ and
$\delta_c+d\delta_c$. This probability is identical to that calculated
directly from the Brownian random walks in section~\ref{sec:der1}.

\section{Comparison with results from Monte-Carlo simulation} \label{sec:mc}

\begin{figure}
  \centering \resizebox{7cm}{9.5cm}{ \includegraphics{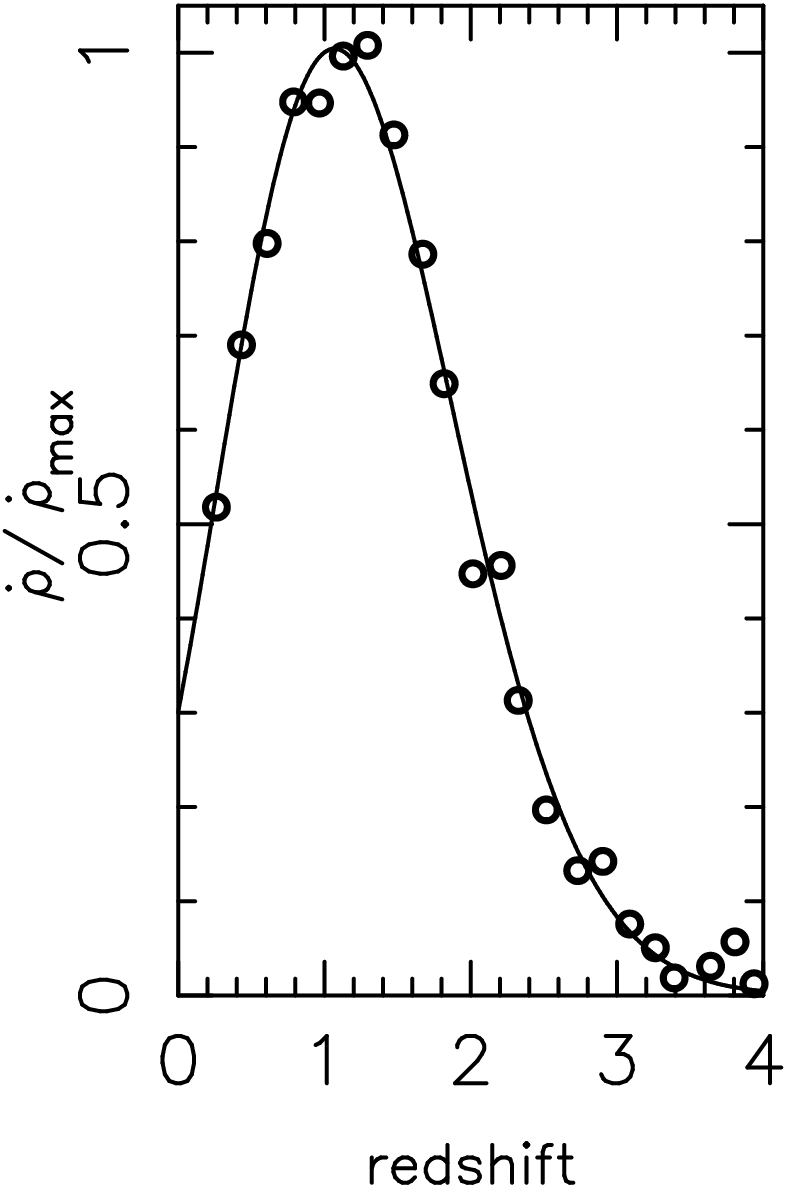} }
  \caption{Comparison of Monte-Carlo results with Press-Schechter
  predictions of the halo formation rate at fixed final mass. Open
  symbols show Monte-Carlo results of the formation time of a halo of
  mass $1.3\times10^{13}\msun$ for a standard CDM power spectrum with
  shape parameter $\Gamma=0.25$ normalised to $\sigma_{8}=0.64$. The
  solid curve shows the prediction of equation~\ref{eq:simple} at this
  mass. The results are normalised to give a peak formation rate of
  1.}
\label{fig:mc}
\end{figure}

As a check on the validity of equation~\ref{eq:simple} we have
performed Monte-Carlo simulations to calculate random trajectories and
determine the distribution of formation epochs.  Approximately
$10^{6}$ random walks were constructed, each consisting of $2^{11}$
uniform steps in $\sigma_M^2$ between 0 and $\sigma_M^2$ equivalent to
a mass of $1.3\times10^{13}\msun$ for a standard CDM power spectrum
with shape parameter $\Gamma=0.25$ normalised to $\sigma_{8}=0.64$. Of
these walks, $10^{4}$ were recorded with first upcrossing points at
the final value of $\sigma_M^2$. The distribution of these upcrossing
points was recorded in uniform bins in time and is compared with the
expected distribution given by equation~\ref{eq:simple} in
Fig~\ref{fig:mc}. Good agreement is demonstrated.

\section{Comparison with results from N-body simulation} 

\begin{figure}
  \centering \resizebox{7cm}{9.5cm}{ \includegraphics{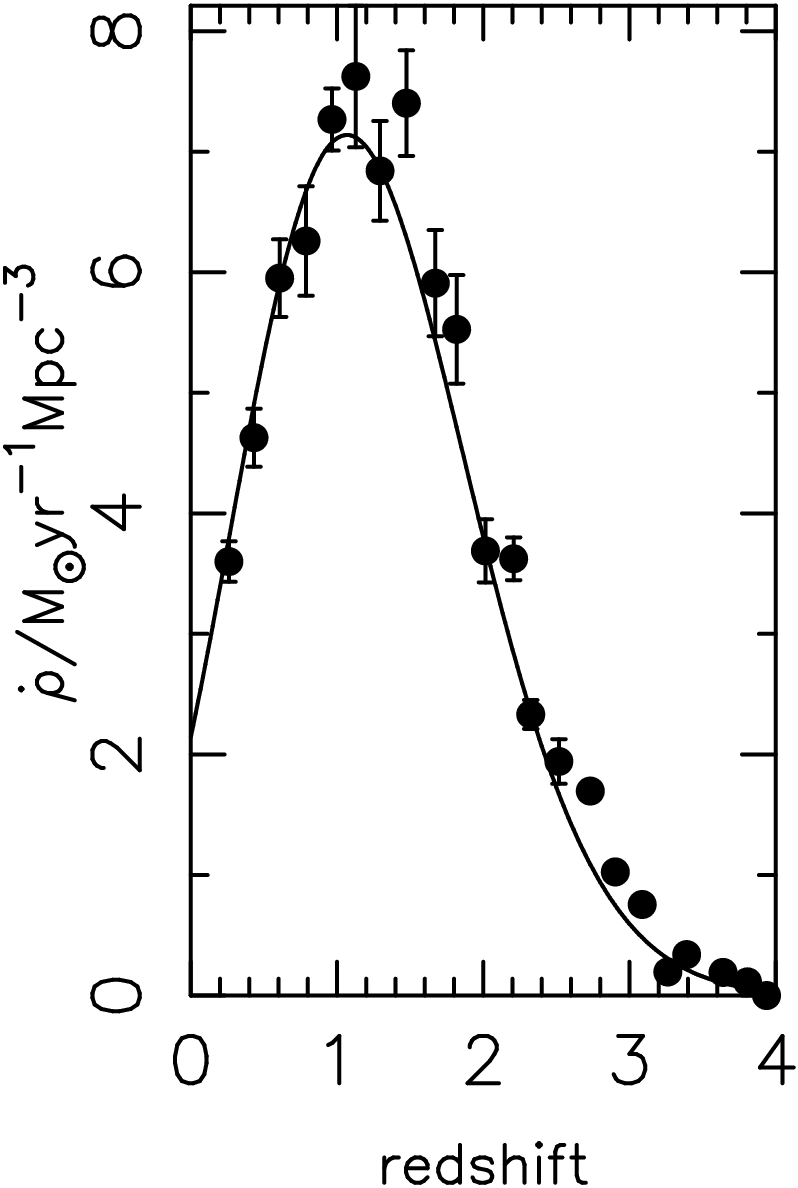} }
  \caption{Comparison of N-body results with Press-Schechter
  predictions of the halo formation rate at fixed final mass. The solid
  symbols show the N-body results for groups of between 45 and 55
  particles, corresponding to a mass of approximately
  $1.3\times10^{13}\msun$, see text for details. The solid curve shows
  the prediction of equation~\ref{eq:simple} at this mass with
  parameters as given in the text and normalised to the N-body
  values.}
\label{fig:nbody}
\end{figure}

We present results from a simulation run using the Hydra N-body,
hydrodynamics code \cite{couchman}. The simulation used $128^{3}$
dark matter particles in a flat universe with $\Omega_{M}=1$,
$\Omega_{\Lambda}=0$, $h=0.5$, and standard CDM power spectrum with
shape parameter $\Gamma=0.25$. The power spectrum was normalised to
$\sigma_{8}=0.64$. Groups of particles were found using a standard
friends of friends (FOF) algorithm which links two particles together
if the mean overdensity of a grid of particles with the same
separation is greater than that predicted at the moment of
virialisation for spherical top-hat collapse \cite{peebles}.

In order to calculate the rate at which halos form we analysed results
from the N-body run at 104 different times, separated by equal
intervals of time. By comparing the FOF results at each output time to
results from earlier times we were able to build up a picture of the
hierarchical growth of structure. Halos with at least half of the
component particles not observed in a halo of equal or higher mass at
an earlier time were recorded as being formed in the time interval
between this output time and the previous one. There is a problem in
this analysis that we may miss formation events - a halo formed in the
time interval of interest may have already merged into a halo of
higher mass when we analysed the simulation. However we can estimate
the potential error caused by this effect by calculating the maximum
amount of halos which could have been formed from the data on
\emph{all} new halos. The distribution of new halos of between 45 and
55 particles is shown in Fig.~\ref{fig:nbody}. The symbols mark the
average of the minimum and maximum mass which could have been involved
in mergers per unit time. The minimum is just the recorded mass in new
halos of between 45 and 55 particles, the maximum is the mass in all
new halos of greater than 45 particles which could have passed through
the interval of interest. The error bars mark the positions of these
maximum and minimum points. Note that these error bars only denote the
error in measuring the mass formation rate from the N-body simulation
results - they do not include errors intrinsic to the N-body
simulation.

In order to count enough formation events we had to use a reasonably
large interval in mass, corresponding to between 45 and 55
particles. However, the curve predicted from equation~\ref{eq:simple}
is similar within this mass range, and the error caused by this effect
will be small. In Fig.~\ref{fig:nbody} we plot the prediction from
equation~\ref{eq:simple} at a mass equivalent to 50 particles which
shows remarkably good agreement with the results of the N-body
simulation. Note that we have not renormalised any of the parameters
of the PS models other than by the self-consistent method given in
section~\ref{sec:ps}.  One caveat we should note, however, is that we
have only been able to test the agreement between the PS result and
N-body simulation at relatively high halo masses.

\section{Comparison with previous results}

We consider that the formation of a halo occurs when all of the mass
in the halo is assembled. Blain \& Longair \shortcite{blain} and
Sasaki \shortcite{sasaki} also used the same definition of
`formation', and tried to calculate the rate of halo formation from
the standard PS mass function (equation~\ref{eq:ps}) and its
derivative. However, they both made assumptions beyond standard PS
theory. Blain \& Longair \shortcite{blain} calculated the formation
rate empirically assuming a form for the distribution of mergers which
occur. Sasaki \shortcite{sasaki} obtained an equation for the
formation rate assuming the destruction rate (from which they then
calculate the formation rate) for a power-law spectrum has no
characteristic mass scale. These assumptions lead to different forms
for the formation rate from that derived above where we have not made
any assumptions beyond those of standard PS theory.

Lacey \& Cole (1993) defined the formation time as the time when the
largest progenitor of a halo first contains at least half the mass of
the halo. Using our definition of formation, we see that the
distribution of such times is equivalent to the formation time
distribution of progenitors of mass greater than M/2 given that they
are the first such progenitors for a particular halo. The formation
time distribution of progenitors of mass M given by
equation~\ref{eq:condt} is {\em only} dependent on knowing that a
progenitor of mass M is formed at some epoch. For instance by setting
$M=M'/2$ in equation~\ref{eq:condt} we can calculate the distribution
of times at which progenitors of mass $M'/2$ form, given that a
progenitor of mass $M'/2$ is formed at some time. However, we do not
know that the first subclump has mass $M'/2$, only that it lies in the
range $M'/2<M<M'$ - this is a consequence of PS theory containing mass
jumps. Lacey \& Cole (1993) give a counting argument (their section
2.5.2) which converts from a probability density in mass to a
distribution in the number of progenitors in order to set the
condition that we are only interested in the {\em first} halo to
contain at least half the mass of the final halo. Without such an
argument it is difficult to see how to distinguish the first
progenitor from subsequent ones and so the results of 
this paper are not directly comparable to those of Lacey \& Cole.

\begin{table}
  \centering
  \begin{tabular}{|c|c|c|c|c|c|} \hline
    model & $\Omega_{M}$ & $\Omega_{\Lambda}$ & $\Gamma$ & $h$ &
    $\sigma_{8}$ \\ \hline OCDM & 0.3 & 0 & 0.15 & 0.5 & 0.85 \\
    $\Lambda$CDM & 0.3 & 0.7 & 0.15 & 0.5 & 0.91 \\ SCDM & 1 & 0 & 0.5 &
    0.5 & 0.60 \\ $\Gamma$CDM & 1 & 0 & 0.25 & 0.5 & 0.60 \\ \hline
  \end{tabular}
  \caption{Table of parameters used for each of the 4 cosmological
    models chosen. Here $h=H_{0}/100$\,km\,s$^{-1}$\,Mpc$^{-1}$,
    $\Omega_{M}\equiv\frac{8\pi G}{3H_{0}^{2}}\rho_{M0}$,
    $\Omega_{\Lambda}\equiv\frac{\Lambda}{3H_{0}^{2}}$, and $\Gamma$ is
    the CDM power spectrum shape parameter.
  }
  \label{tab:models} 
\end{table}

\section{The predicted halo formation rate}

We can now estimate the rate at which mergers occur to create halos of
any particular mass for a number of cosmologies. We investigate 4
different cosmologies, based on those used by the Virgo consortium
\cite{springel,jenkins} summarised in Table~\ref{tab:models}.

\begin{figure}
  \centering
  \resizebox{7cm}{9.5cm}{
    \includegraphics{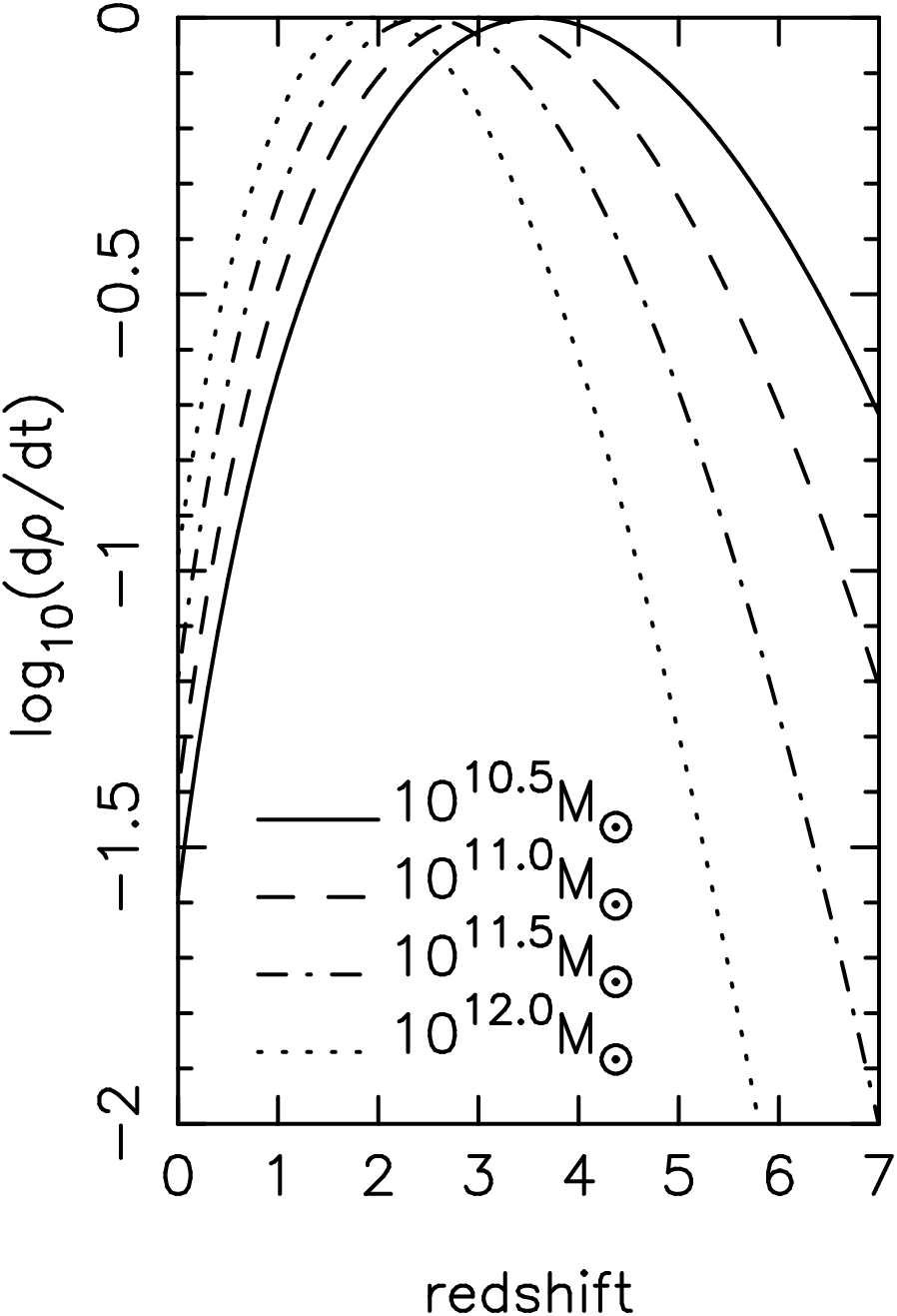}
  }
  \caption{The formation rate of galaxy halos as calculated
  using equation~\ref{eq:rate_cstm} for the $\Gamma$CDM cosmological
  model. Curves are plotted for halo masses in the range
  $10^{10.5} - 10^{12.0} \msun$ and are normalised to give the same peak
  formation rate. } 
\label{fig:model_one}
\end{figure}

Fig.~\ref{fig:model_one} shows how the functions vary with halo mass
for the $\Gamma$CDM model. Curves are plotted at 4 different values of
halo mass, $10^{10.5} - 10^{12}\msun$. Here it is easy to see the
hierarchical build up of structure - low mass halos predominantly form
first, with their peak formation rate occurring at higher redshift
than that for higher mass halos.

\begin{figure}
  \centering
  \resizebox{7cm}{9.5cm}{
    \includegraphics{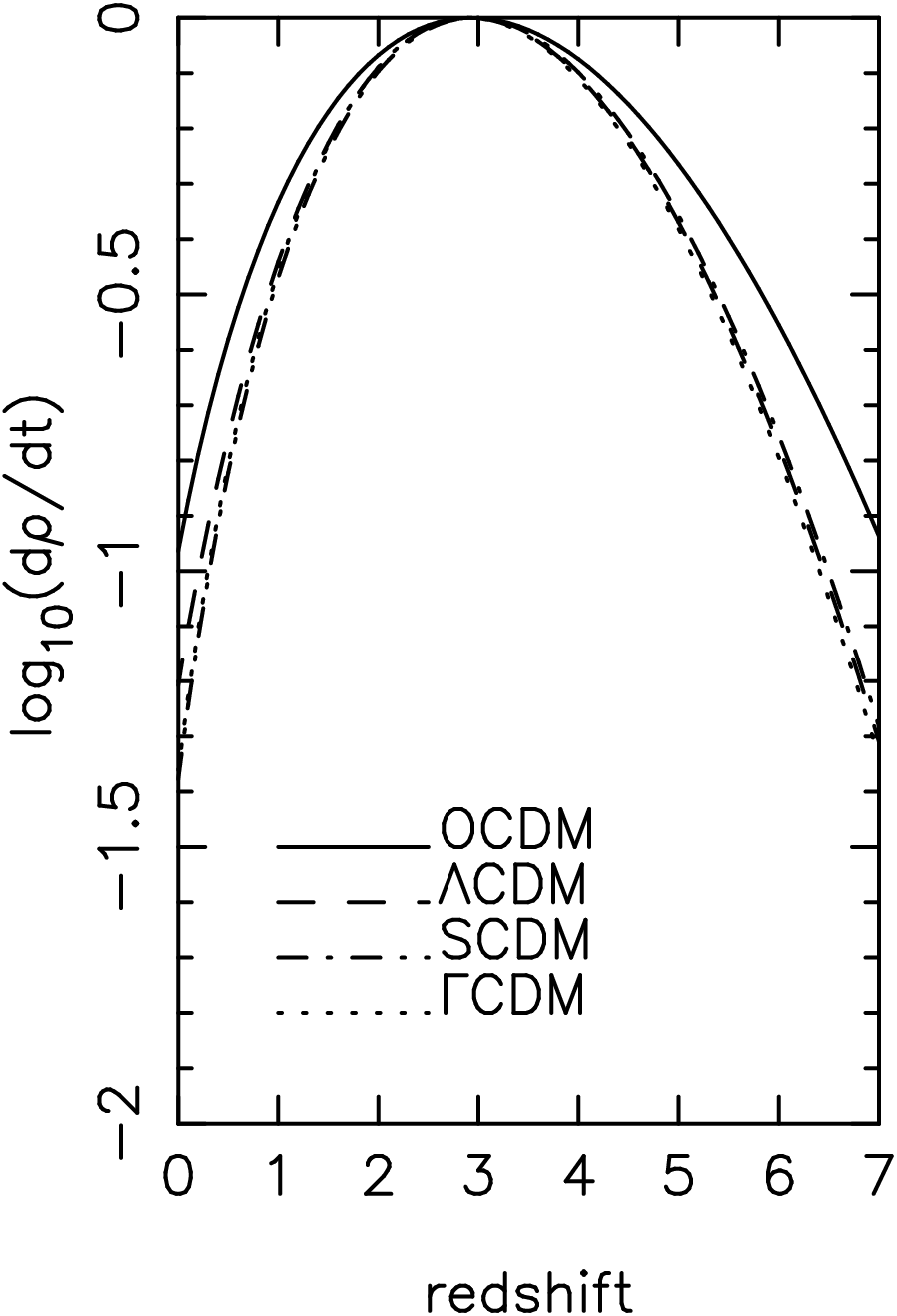}
  }
  \caption{Comparison of the 4 different cosmological models. 
  Because of the differing normalisations of the models the
  formation rates are plotted for different values 
  of the halo mass, chosen to provide the closest match
  between the models. The masses are: OCDM, $10^{12.7}\msun$; 
  $\Lambda$CDM, $10^{12.0}\msun$; SCDM, $10^{12.1}\msun$; 
  $\Gamma$CDM, $10^{11.1}\msun$. 
  The curves are normalised to the same peak
  formation rate. }
\label{fig:conv1}
\end{figure}

For each cosmology chosen, for a particular mass of halo, the redshift
at which the formation rate peaks is highly dependent on the
$\sigma_8$ normalisation chosen. High values of $\sigma_8$ lead to
early formation and the peak in the formation rate is at higher
redshift. However by choosing a higher halo mass we can recover the
same shape of curve. This is easy to see in the SCDM and $\Gamma$CDM
cosmologies, where equation~\ref{eq:simple} applies, and the curve
shape is only dependent on the value of $\beta$ which can be kept
constant by varying the halo mass with $\sigma_8$.
Fig.~\ref{fig:conv1} shows the mass merging rate for each of the
cosmological models, but with halo masses chosen such that the models
approximately agree on the redshift of the peak of halo
formation. Viewed in this way, the OCDM model produces less evolution
than the other models, but the other three have rather similar amounts
of evolution.

\section{The Merger-Induced Star Formation Rate} \label{sec:sfr}

A complete description of the formation of stars in
hierarchically-forming galaxies requires us to understand mergers
between dark halos, the fate of the baryonic matter in such mergers,
and the processes of cooling, star formation and feedback and stellar
evolution that all affect the observed luminosity of a galaxy and its
time variation.  Some recent attempts to explain specifically the
strong evolution in star formation have used {\it ad hoc} models of
the rate of generation of new galaxies as a source function, which
have then been combined with models describing the remaining processes
and tested against the observed data in sub-mm, infrared, optical and
ultraviolet wavebands (Madau et al.\ 1996, 1998).  An alternative
method is to simulate the merger histories of galaxies based on the
work of Lacey \& Cole (1993, 1994), add heuristic recipes for the
formation of stars and predict quantities such as the galaxy
luminosity function or the cosmic evolution in star formation rate
\cite{kauffmann,cole94,baugh}.  The success of these latter models has
been the broad general agreement between the observations and the
models, in that evolution in the predicted star formation rate is
seen, with a maximum rate at redshifts about unity (Madau et al.\
1996, 1998; Baugh et al.\ 1998). But when considered in more detail,
it is clear that those models do not reproduce the {\it amount} of
evolution that is seen.  That amount appears to be at least a factor
of about ten (Lilly et al.\ 1996; Connolly et al.\ 1997; Madau et al.\
1996, 1998), and the sub-mm data suggest that factors of about 30 may
be required \cite{hughes}. Such large evolution in star formation rate
would be consistent with the depletion of neutral gas that is inferred
from the evolution in intervening absorption towards distant quasars
\cite{peifall}.  The models cited above appear to predict evolution
only by a factor about 5.  Semi-analytic models incorporating an
additional evolving component of star-formation associated with
short-lived starbursts have now been produced \cite{guiderdoni} in
which cosmological evolution by a factor $(1+z)^{5}$ is assumed, and
these models do appear to give a better representation of the observed
SFR.

Our aim in this section is to provide a complementary analysis to the
above work by considering only the role of the cosmological evolution
in halo formation rate.  Observed luminous starbursts at low redshifts
are probably rather young, and simulations indicate that bursts of
star formation associated with mergers between either disk or
bulge/halo systems with a variety of relative masses all have
lifetimes which are short ($<10^8$\,years) (Mihos \& Hernquist 1994,
1996).  We have already argued that the rate of halo formation should
have the same cosmological evolution as the rate of mergers between
dark halos, and we should therefore expect any star-formation which is
associated with mergers to have cosmological evolution consistent with
the calculated evolution in halo formation rate.  Similarity between
the observed and calculated evolution would indicate that dark halo
mergers are an important factor in star formation at high redshifts
and we would then predict that most of the star formation in high
redshift galaxies should be short-lived.

\begin{figure}
  \centering \resizebox{7cm}{9.5cm}{\includegraphics{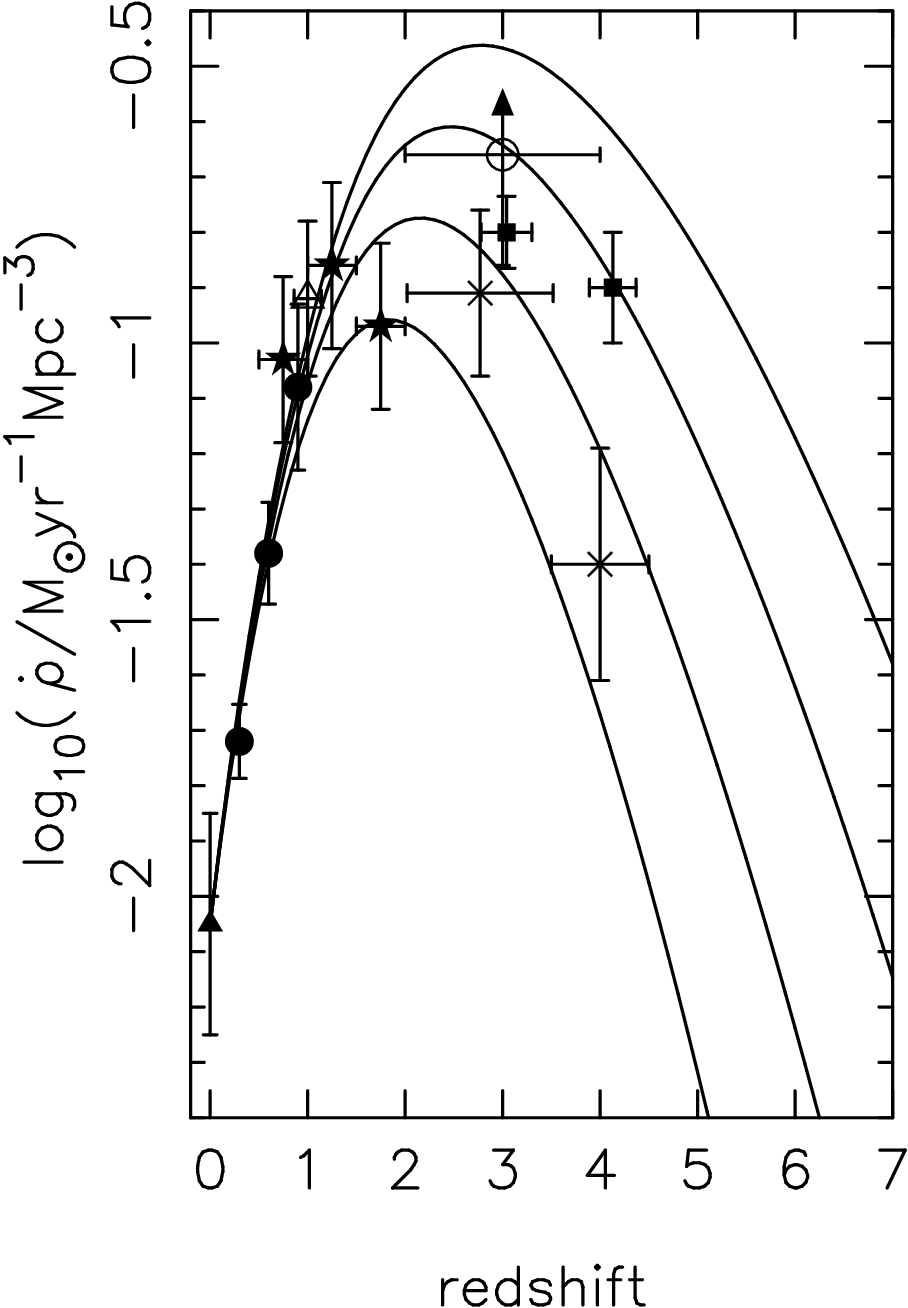}}
  \caption{The observed mean comoving volume-averaged star formation
  rate as determined from the Canada-France redshift survey (Lilly et
  al.\ 1996; Madau et al.\ 1996) (solid circles), optical HDF data
  corrected for dust extinction (Madau et al.\ 1996; Pettini et al.\
  1997) (crosses) and Connolly et al. (1997) (solid stars), 
  extinction-corrected Lyman
  break galaxies (Steidel et al.\ 1998c) (solid squares), sub-mm data
  (Hughes et al.\ 1998) (open circle) and the star formation rate
  inferred from H$\alpha$ surveys at zero-redshift (Gallego et al.\
  1995) (solid triangle) and at redshift 1 (Glazebrook et al.\ 1998)
  (open triangle). A Salpeter IMF and flat $\Omega_{M}=1$ cosmology
  has been assumed. The
  curves are the predictions of the $\Gamma$CDM model  
  for masses of $10^{10.0}, 10^{10.5}, 10^{11.0},
  10^{11.5} \msun$ normalised to the local star formation rate (see
  Section \ref{sec:sfr}).}
\label{fig:data2}
\end{figure}

The inferred star formation rates shown in Fig.~\ref{fig:data2} are
derived primarily from the observed luminosity density at a rest-frame
wavelength of 280\,nm.  Models of an evolving galaxy after a burst of
star formation \cite{bruzual} show that the ultraviolet light decays
approximately exponentially with a timescale of about 0.6\,Gyr and it
is this timescale which therefore dominates the light curve of
short-lived starbursts.  Thus, to compare the cosmic evolution in
formation rate with the observed evolution we convolve the rate of
halo formation as derived above with an exponential function with this
decay timescale.  Note that this simple approach is not designed to
provide a complete description of the physics of star formation in
these galaxies: merely to test the relative importance of the cosmic
variation in merger rate.

A sample of such curves are plotted with recent star formation rate
data in Fig.~\ref{fig:data2}: to be consistent with this
previously-published data we adopt the $\Omega_{M}=1$,
$\Omega_{\Lambda}=0$ cosmology and we show a range of halo masses for
the $\Gamma$CDM model. Since the fraction of halo mergers which result
in an observable starburst is unknown, we treat the normalisation of
the models as being a free parameter and normalise the curves to the
zero-redshift data-point.

\begin{figure}
  \centering \resizebox{7cm}{9.5cm}{\includegraphics{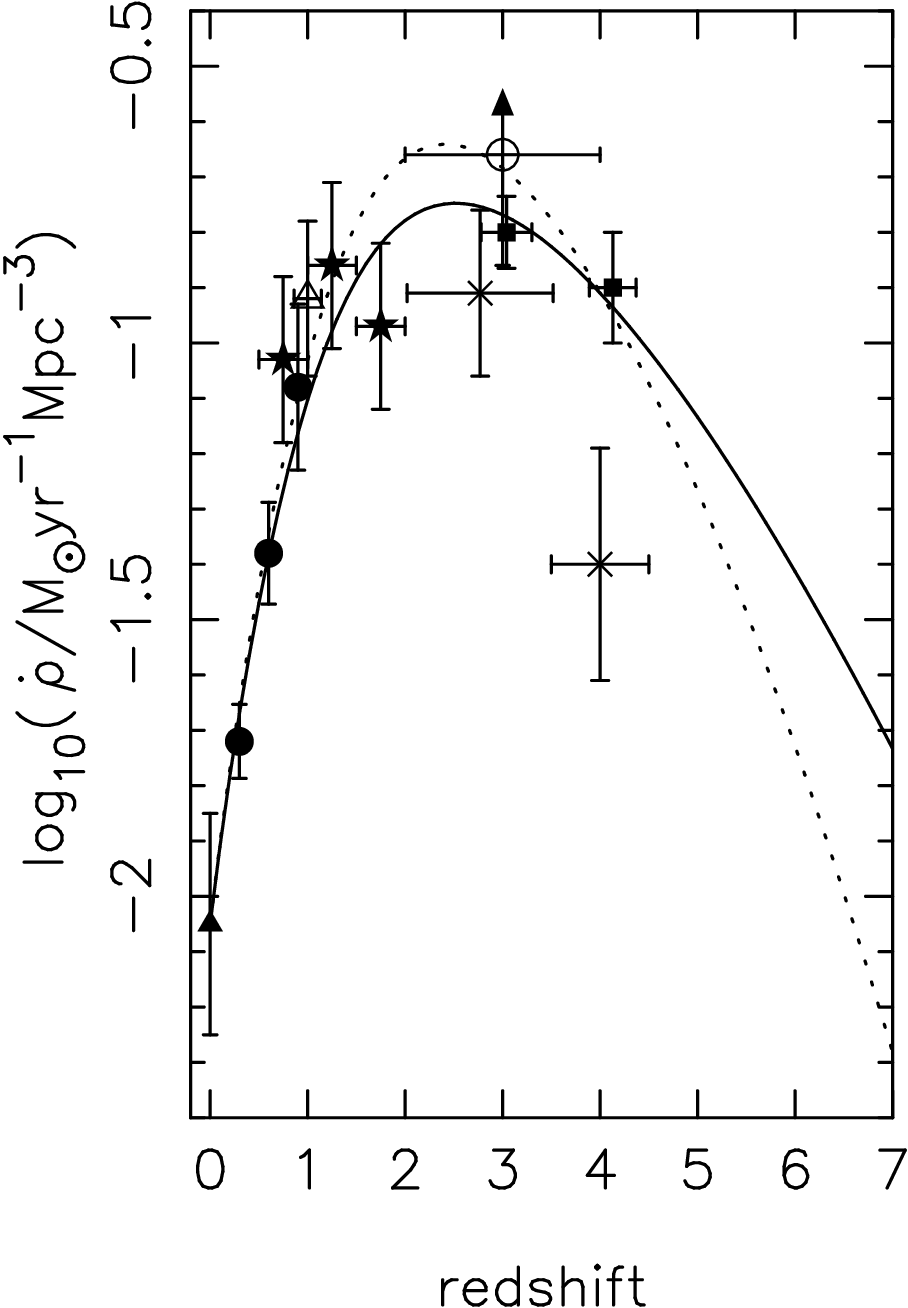}}
  \caption{ The evolution predicted by combining the formation rate of
  different mass halos (convolved with an expected lifetime of 0.6Gyr)
  weighted by a Gaussian in dn/dlogM (solid line). The Gaussian
  is centred on a halo mass of $10^{10.6}\msun$ and has
  $\sigma=1.5$. The evolution expected from halo
  formation events resulting in halos only of mass $10^{10.6}\msun$ is
  also plotted (dotted line).  Both curves are normalised to the data
  at $z=0$.  Data are plotted as in Fig.~\ref{fig:data2}.}
\label{fig:sfr}
\end{figure}

We see excellent agreement between the data and the cosmic variation
in halo formation rate for any of the dark halo masses considered at
redshifts up to about 1.  At higher redshifts the data are very
uncertain, and will remain so until corrections for optical extinction
in starbursts are better understood, but we expect the relative
contributions from a range of halo masses to be important in
influencing the observed form of the evolution at high redshift. As an
illustration, we combine formation rates for halos of a range of final
masses, weighting the contribution from each mass by a Gaussian
distribution in dn/dlogM. The relative normalisation for each mass is
determined at $z=0$ using equation~\ref{eq:ps}.  Fig.~\ref{fig:sfr}
shows the resulting evolution compared with the star formation data,
for a Gaussian weighting function centred on $10^{10.6}\msun$ with a
logarithmic dispersion $\sigma=1.5$.  Combining the curves in this way
flattens the high redshift evolution as mergers creating lower mass
halos become increasingly important. As expected, the $0<z<1$
evolution is the same as that shown in Fig~\ref{fig:data2}, but there
is now much better agreement with the high-redshift data. A thorough
treatment of this issue is beyond the scope of this paper, and would
require additional modelling such as may be obtained from the
semi-analytic approach. The illustration presented here is sufficient
to show that the observed high-redshift behaviour can be reproduced by
considering the star-forming galaxies to be dominated by short-lived
merger-induced starbursts, and that it is the cosmic variation in halo
formation rate that drives the evolution in star formation.

\section{Quasar Density Evolution} \label{sec:qde}

\begin{figure}
  \centering \resizebox{7cm}{9.5cm}{\includegraphics{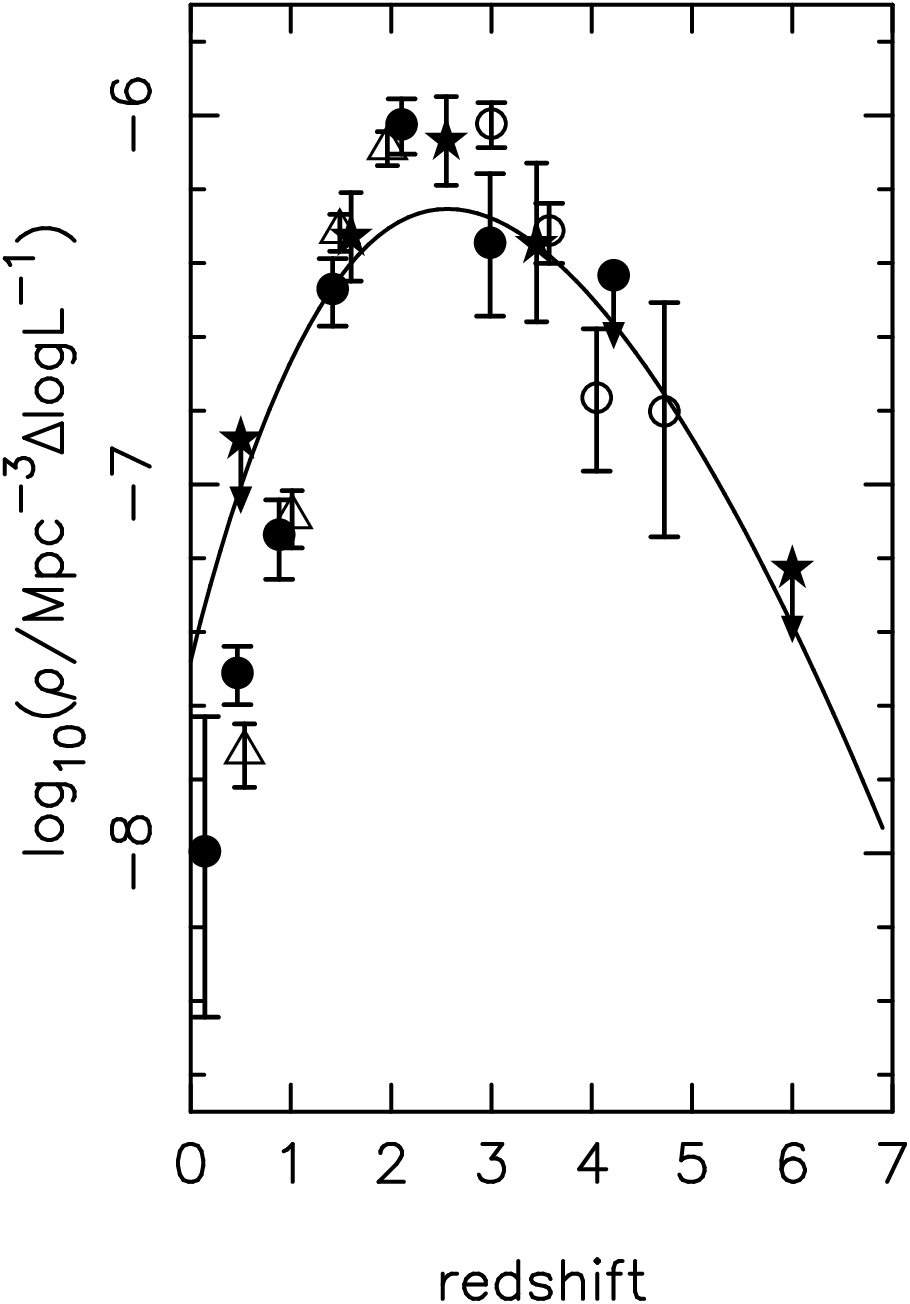}}
  \caption{ The measured comoving space density of quasars is plotted
  for flat-spectrum radio-loud quasars(Shaver et al.\ 1996) with
  $P_{2.7 GHz} > 10^{27.04}$\,Watt\,Hz$^{-1}$\,sr$^{-1}$
  (\emph{stars}), optically-selected UVX quasars(Goldschmidt \& Miller
  1998; Boyle et al.\ 1990) with $-25.4 > M_B > -27.9$ (\emph{open
  triangles}), optically-selected high-redshift quasars(Schneider,
  Schmidt \& Gunn 1994) with $10^{38} < P_{Ly-\alpha} {\rm Watt} <
  10^{39}$ (\emph{open circles}), and a composite of ROSAT soft X-ray
  selected quasars(Boyle, Wilkes \& Elvis 1998; Boyle et al.\ 1994;
  Read et al. 1999; Schmidt et al.\ 1998) with $10^{37.7} < P_{0.5-2 keV}
  Watt < 10^{38.7}$ (\emph{filled circles}).  The cosmological
  parameters of the $\Lambda$CDM model have been assumed: the shape of
  the evolution shows little variation between the three cosmologies
  parameterised in Table\,1.  As each waveband selects a different
  subset of the overall quasar population, and also covers a slightly
  different range of bolometric luminosity, the normalisation of each
  of the samples has been adjusted such that they lie on a consistent
  curve passing through all the data and then normalised to the X-ray
  data.  The curve plotted is the predictions of the $\Lambda$CDM
  model for a halo mass of $10^{11.8} \msun$ with normalisation of the
  curve adjusted to match the data (see Section \ref{sec:qde}). }
\label{fig:data1}
\end{figure}

The comoving space density of quasars has also decreased by a factor
of order 100 from a redshift of 2 to the present day \cite{boyle}.
The quasar comoving space density appears to have a maximum at a
redshift about 2--3 \cite{ssg,shaver}. Fig.~\ref{fig:data1} shows a
compilation of measurements of the comoving space density of quasars
selected in one decade of luminosity, shown as a function of redshift.
The compilation shows quasars selected in radio, optical and X-ray
wavebands: the consistency in their evolution is a strong argument
that the form of the observed evolution is intrinsic to the quasar
population and not due to either observational selection effects or to
intervening obscuration (see also Shaver et al.\ 1996).  The
luminosity range that has been chosen is that in which the greatest
amount of cosmological evolution is seen in the range $0 < z < 2$, and
this therefore provides the most stringent test of the model we
discuss here.  Previous authors have chosen instead to plot the
comoving quasar luminosity density integrated over a wide range in
luminosity (e.g. Boyle \& Terlevich 1998): this produces a lower
amount of evolution, in fact more comparable to the inferred evolution
in SFR.  We adopt the more pessimistic approach here.

Our knowledge of the physics of quasar activation is even worse than
our knowledge of the activation of intense bursts of star
formation. We follow on from work by Efstathiou \& Rees
\shortcite{efrees} and Haehnelt \& Rees \shortcite{haehnelt} and
assume that merger events activate quasars, although we are not
concerned with the actual mechanism of the activation. As for the
induced starbursts we wish to establish whether the cosmic evolution
in halo formation rate can account for much of the evolution in quasar
numbers.

We have little knowledge of the lifetime of quasars, but we can see
that their lifetime must be short if the model is to reproduce the
observed amount of evolution.  The association of a few quasars with
starbursts might be taken as evidence that their lifetimes are
similar, and the ages of extended radio sources are also thought to be
short ($< 10^8$\,years).  We have therefore plotted the formation rate
convolved with the same exponential lifetime as used to model
starburst evolution with the quasar data in Fig.~\ref{fig:data1}. The
normalisation has again been treated as a free parameter.  Recent
observations of luminous quasars suggest that they live in massive
host galaxies with a narrow range of host luminosities (McLeod \&
Rieke 1994a,b, 1995; Taylor et al.\ 1996; Hooper, Impey \& Foltz 1997)
and by inference a narrow range of host masses.  Here we compare the
model at a single halo mass to the data. In this case the OCDM model
does not produce enough evolution for any mass of halo and does not
fit the data well.  Reasonable fits can be obtained for the remaining
three models, although all of them have a deficit in the amount of
evolution between $z=2$ and $z=0$ of about a factor 4.  We have chosen
to plot the $\Lambda$CDM model as, with its high $\sigma_8$, halos of
masses $10^{11.8} \msun$ peak at $z \sim 2.5$.  The SCDM model also
fits reasonably with a high value for halo mass, but can probably be
excluded on other grounds \cite{gawiser}.  The $\Gamma$CDM model fits
with a value for the halo mass of $10^{10.6} \msun$: a value which is
rather lower than would be implied by the observation that luminous
quasars exist in massive host galaxies. The halo mass would be
increased to $10^{11.8} \msun$ if $\sigma_8$ were increased to 0.9:
such high power-spectrum normalisations are indicated by the COBE data
\cite{gawiser}.  Finally, we note that increasing $h$ from 0.5 to 0.7
decreases the inferred halo masses by a factor about 3.

\section{Conclusions}

We have derived a simple formula for the rate at which structure
merges to form new halos within Press-Schechter theory. This formula
has been tested and shown to be in good agreement with the results of
numerical simulations.

We have also considered whether the strong cosmological evolution in
star-formation is primarily driven by the cosmic variation in the rate
of halo formation.  In order to produce a complete model of the
evolution of merger-induced star formation we need to know much more
physics than we have used in this paper. However, it is expected that
the evolution in the rate at which dark matter halos merge to form
higher mass halos will have a large influence on the evolution of the
merger-induced SFR. Examining Fig.~\ref{fig:data2} we saw that,
although it is expected that a range of halo masses are important in
producing star-bursts, for halo masses $10^{10}-10^{11.5}\msun$ the
observed drop in the SFR from redshifts $z\sim1$ to $z\sim0$ is
independent of mass and the evolution predicted from any linear
combination of these curves shows strong evolution. The similarity of
this evolution to that observed is remarkable and, given that
quiescent star formation predicted from semi-analytic models does not
provide enough evolution in this redshift range \cite{guiderdoni},
indicates that merger-induced starbursts are extremely important for
star formation at $z \sim 1$ and are perhaps the principal mechanism
leading to the observed star formation at high redshifts. This model
does not preclude the undoubted existence of quiescent star-formation
as well: it merely states that because the starburst component evolves
so strongly it dominates at high redshift.  At higher redshifts a more
physically-motivated model is needed to deduce the relative
contributions of a range of halo masses, but we have shown that a
simple combination of such a range can produce evolution which is in
good agreement with the data of Steidel et al. \shortcite{steidel98}.

The evolution of the quasar population at optical absolute magnitudes
$M_B \sim -26$ is larger than that of the star formation rate,
although the evolution of the total quasar luminosity density is not
\cite{bt}.  We have shown that, provided that quasar lifetimes are
finite but short (around 0.6\,Gyr), then mergers to form the halos of
massive galaxies can reproduce the principal features of the observed
quasar evolution, consistent with the observation that luminous
quasars consistently live in massive host galaxies.  But whilst the
curve predicts very well all the basic features of the observations,
it fails by about a factor of $\sim$ 4 to produce enough evolution
between redshifts of 0 and 2.  In this sense the model does not
provide a full explanation of the observed evolution, but as the PS
theory does not include any non-linear dynamical effects we consider
even this agreement to be surprisingly good.  Previous work has
suggested that the expected cosmic variation in galaxy velocity
dispersion \cite{carlberg} or in halo circular velocity
\cite{haehnelt} could contribute to the quasar evolution, and only a
modest amount of additional evolution would be needed.  In the latter
case we should expect to see some cosmological evolution in the host
masses associated with a given quasar luminosity.  Some additional
component of quasar luminosity evolution \cite{boyle,goldschmidt}
would also have the desired effect.  Quasar lifetimes significantly
longer than about 1\,Gyr would smooth out the evolution in these
models and would produce worse agreement with the data.

From the above comparisons between observations and the calculated
rate of dark halo formation we conclude that hierarchical merging is
highly important for the strong cosmological evolution that is
observed in both star formation and quasar activity. Determination of
the masses of halos involved in star formation at high redshifts
(e.g. Pettini et al. 1998) and quasars would provide good constraints
on allowed cosmological models and their parameters, and estimation of
the ages of the star-forming systems at high redshift would be a
direct test of the proposition that short-lived starbursts account for
much of the observed star formation.

\section{Acknowledgements}
We are grateful for the use of the Hydra N-body code \cite{couchman}
kindly provided by the Hydra consortium. WJP acknowledges a PPARC
studentship.


\begin{thebibliography}{}
  \bibitem[\protect\citename{Adelberger et al.\ }1998]{adelberger} 
    Adelberger K., Steidel C., Giavalisco M., Dickinson M., Pettini
    M., Kellogg M., 1998, ApJ, 505, 18. 
  \bibitem[\protect\citename{Barnes \& Hernquist }1992]{barnes92} 
    Barnes J.E., Hernquist L., 1992, ARA\&A, 30, 705
  \bibitem[\protect\citename{Baugh et al.\ }1998]{baugh} 
    Baugh C.M., Cole S., Frenk C.S., Lacey C.G., 1998, ApJ, 498, 504
  \bibitem[\protect\citename{Blain \& Longair }1993]{blain} 
    Blain A.W., Longair M.S., 1993, MNRAS, 264, 509
  \bibitem[\protect\citename{Bond et al.\ }1991]{bond} 
    Bond J.R., Cole S., Efstathiou G., Kaiser N., 1991, ApJ, 379, 440
  \bibitem[\protect\citename{Boyle et al.\ }1990]{boyle} 
    Boyle B.J., Fong R., Shanks T., Peterson B.A., 1990, MNRAS, 243, 1
  \bibitem[\protect\citename{Boyle et al.\ }1994]{bsgsg} 
    Boyle B.J., Shanks T., Georgantopoulos I., 
    Stewart G.C., Griffiths, R.E., 1994, MNRAS, 271, 639
  \bibitem[\protect\citename{Boyle, Wilkes \& Elvis }1998]{bwe} 
    Boyle B.J., Wilkes B.J., Elvis M., 1998, MNRAS, 285, 511
  \bibitem[\protect\citename{Boyle \& Terlevich }1998]{bt} 
    Boyle B.J., Terlevich R.J., 1998, MNRAS, 293, L49.
  \bibitem[\protect\citename{Brotherton et al }1999]{brotherton}
    Brotherton M.S. et al., 1999, in preparation.
  \bibitem[\protect\citename{Bruzual \& Charlot }1993]{bruzual} 
    Bruzual A. G., Charlot S., 1993, ApJ, 405, 538
  \bibitem[\protect\citename{Canalizo \& Stockton }1997]{canalizo} 
    Canalizo G., Stockton A., 1997, ApJ, 480, L5
  \bibitem[\protect\citename{Carlberg }1990]{carlberg} 
    Carlberg R.G., 1990, ApJ, 350, 505
  \bibitem[\protect\citename{Cole et al.\ }1994]{cole94} 
    Cole S., Aragon-Salamanca A., Frenk C.S., 
    Navarro J.F., Zepf S.E., 1994, MNRAS, 271, 781
  \bibitem[\protect\citename{Connolly et al.\ }1997]{connolly} 
    Connolly A.J., Szalay A.S., Dickinson M.,
    SubbaRao M.V., Brunner R.J., 1997, ApJ, 486, L11
  \bibitem[\protect\citename{Couchman, Thomas \& Pearce }1995]{couchman} 
    Couchman H. M. P., Thomas P.A., Pearce F.R., 1995, ApJ, 452, 797
  \bibitem[\protect\citename{Dunlop }1997]{dunlop} 
    Dunlop J., 1997, `Observational Cosmology with the New
    Radio Surveys', eds. Bremer et al., Kluwer 
  \bibitem[\protect\citename{Efstathiou \& Rees }1988]{efrees} 
    Efstathiou G., Rees M.J., 1988, MNRAS, 230, 5p
  \bibitem[\protect\citename{Efstathiou et al.\ }1988]{efstathiou} 
    Efstathiou G., Frenk C.S., White S.D.M., Davis M., 1988, MNRAS, 235, 715
  \bibitem[\protect\citename{Eke, Cole \& Frenk }1996]{eke} 
    Eke V.R., Cole S., Frenk C.S., 1996, MNRAS, 282, 263
  \bibitem[\protect\citename{Gallego et al.\ }1995]{gallego} 
    Gallego J., Zamorano J., Aragon-Salamanca A., Rego M., 1995, ApJ, 455, L1
  \bibitem[\protect\citename{Gawiser \& Silk }1998]{gawiser} 
    Gawiser E., Silk J., 1998, Sci, 280, 1405
  \bibitem[\protect\citename{Glazebrook et al.\ }1998]{glazebrook} 
    Glazebrook K., Blake C., Economou F., Lilly S., Colless M., 1998,
    MNRAS submitted, astro-ph/9808276 
  \bibitem[\protect\citename{Goldschmidt \& Miller }1998]{goldschmidt}
    Goldschmidt P., Miller L., 1998, MNRAS, 293, 107
  \bibitem[\protect\citename{Gunn \& Gott }1972]{gunn} 
    Gunn J.E., Gott J.R., 1972, ApJ, 176, 1
  \bibitem[\protect\citename{Guiderdoni et al.\ }1998]{guiderdoni}
    Guiderdoni B., Hivon E., Bouchet F.R., Maffei B., 1998, MNRAS, 295, 877
  \bibitem[\protect\citename{Haehnelt \& Rees }1993]{haehnelt} 
    Haehnelt M.G., Rees M.J., 1993, MNRAS, 263, 168
  \bibitem[\protect\citename{Hooper, Impey \& Foltz }1997]{hooper} 
    Hooper E.J., Impey C.D., Foltz C.B., 1997, ApJ, 480, L95
  \bibitem[\protect\citename{Hughes et al.\ }1998]{hughes} 
    Hughes D. et al., 1998, Nat, 394, 241
  \bibitem[\protect\citename{Jedamsik }1995]{jedamsik} 
    Jedamsik K., 1995, ApJ, 448, 1
  \bibitem[\protect\citename{Jenkins et al.\ }1998]{jenkins} 
    Jenkins A. et al., 1998, ApJ, 499, 20
  \bibitem[\protect\citename{Karlin \& Taylor }1975]{karlin} 
    Karlin S., Taylor H.M., 1975, A first course in stochastic
    processes, 2nd ed. London Academic Press
  \bibitem[\protect\citename{Kauffmann, White \& Guiderdoni
    }1993]{kauffmann} Kauffmann G., White S.D.M., Guiderdoni B.,
    1993, MNRAS, 264, 201
  \bibitem[\protect\citename{Lacey \& Cole }1993]{lc93} 
    Lacey C., Cole S., 1993, MNRAS, 262, 627  
  \bibitem[\protect\citename{Lacey \& Cole }1994]{lc94} 
    Lacey C., Cole S., 1994, MNRAS, 271, 676
  \bibitem[\protect\citename{Lilly et al.\ }1996]{lilly} 
    Lilly S.J., Le Fevre O., Hammer F., Crampton D., 1996, ApJ, 460, L1
  \bibitem[\protect\citename{Madau et al.\ }1996]{madau} 
    Madau P., Ferguson H.C., Dickinson M.E.,
    Giavalisco M., Steidel C.C., Fruchter A., 1996, MNRAS, 283, 1388
  \bibitem[\protect\citename{Madau, Pozzetti \& Dickinson }1998]{madau2} 
    Madau P., Pozzetti L., Dickinson M., 1998, ApJ, 498, 106
  \bibitem[\protect\citename{Mihos \& Hernquist }1994]{mh1} 
    Mihos J.C., Hernquist L., 1994, ApJ, 425, L13
  \bibitem[\protect\citename{Mihos \& Hernquist }1996]{mh2} 
    Mihos J.C., Hernquist L., 1996, ApJ, 464, 641
  \bibitem[\protect\citename{McLeod \& Rieke }1994a]{mcleoda} 
    McLeod K.K., Rieke G.H., 1994a, ApJ, 420, 58
  \bibitem[\protect\citename{McLeod \& Rieke }1994b]{mcleodb} 
    McLeod K.K., Rieke G.H., 1994b, ApJ, 431, 137
  \bibitem[\protect\citename{McLeod \& Rieke }1995]{mcleodc} 
    McLeod K.K., Rieke G.H., 1995, ApJ, 454, L77
  \bibitem[\protect\citename{Peacock \& Heavens }1990]{peacock} 
    Peacock J. A., Heavens A. F., 1990, MNRAS, 243, 133
  \bibitem[\protect\citename{Peebles }1980]{peebles} 
    Peebles P.J.E., 1980, The large-scale structure of the
    universe.  Princeton University Press
  \bibitem[\protect\citename{Pei \& Fall }1995]{peifall} 
    Pei Y.C., Fall S.M., 1995, ApJ, 454, 69
  \bibitem[\protect\citename{Pettini et al.\ }1997]{pettini} 
    Pettini M., Steidel C.C., Adelberger K.L.,
    Kellogg M., Dickinson M., Giavalisco M., 1997, `Origins',
    Astron. Soc. Pacific Conference Series
  \bibitem[\protect\citename{Pettini et al.\ }1998]{pettini98}
    Pettini M., Kellogg M., Steidel C.C., Dickinson M.,
    Adelberger K.L. \& Giavalisco M., 1998.  ApJ, 508, 539
  \bibitem[\protect\citename{Press \& Schechter }1974]{ps} 
    Press W., Schechter P., 1974, ApJ, 187, 425 
  \bibitem[\protect\citename{Read }1999]{read} 
    Read M.A. et al. in preparation.
  \bibitem[\protect\citename{Sasaki }1994]{sasaki} 
    Sasaki Shin, PASJ, 1994, 46, 427
  \bibitem[\protect\citename{Schmidt, Schneider \& Gunn }1995]{ssg} 
    Schmidt M., Schneider D.P., Gunn J.E., 1995, AJ, 110, 68
  \bibitem[\protect\citename{Schmidt et al.\ }1998]{schmidtrosat} 
    Schmidt M. et al., 1998, A\&A, 329, 495
  \bibitem[\protect\citename{Schneider, Schmidt \& Gunn } 1994]{schneider} 
    Schneider D.P., Schmidt M., Gunn J.E., 1994, AJ, 107, 1245
  \bibitem[\protect\citename{Shaver et al.\ }1996]{shaver} 
    Shaver P.A., Wall J.V., Kellermann K.I., Jackson
    C.A., Hawkins M.R.S., 1996, Nat, 384, 439
  \bibitem[\protect\citename{Somerville et al.\ }1998]{somerville} 
    Somerville R.S., Lemson G., Kolatt T.S., Dekel A., 
    1998, MNRAS in press, astro-ph/9807277	
  \bibitem[\protect\citename{Springel et al.\ }1998]{springel} 
    Springel V. et al., 1998. MNRAS 298, 1169
  \bibitem[\protect\citename{Steidel et al.\ }1998a]{steidel97} 
    Steidel C.C., Adelberger K.L., Dickinson M.,
    Giavalisco M., Pettini M., Kellogg M., 1998a, ApJ, 492, 428
  \bibitem[\protect\citename{Steidel et al.\ }1998b]{steidelrs} 
    Steidel C., Adelberger K., Giavalisco M., Dickinson M., Pettini
    M., Kellogg M., 1998b, Phil Trans. R. Soc. Lond. A, 1750.
  \bibitem[\protect\citename{Steidel et al.\ }1998c]{steidel98} 
    Steidel C., Adelberger K., Giavalisco M., Dickinson M., Pettini
    M., 1998c, ApJ submitted, astro-ph/9811399 
  \bibitem[\protect\citename{Stockton, Canalizo \& Close }1998]{stockton} 
    Stockton A., Canalizo G., Close L.M., 1998, ApJ, 500, L121
  \bibitem[\protect\citename{Taylor et al.\ }1996]{taylor} 
    Taylor G.L., Dunlop J.S., Hughes D.H., Robson E.I., 1996, MNRAS, 283, 930
  \bibitem[\protect\citename{Yano, Nagashima \& Gouda }1996]{yano} 
    Yano T., Nagashima M., Gouda N., 1996, ApJ, 466, 1
\end{thebibliography}
\end{document}